\documentclass{PoS}

\usepackage{graphicx}
\usepackage{amsmath}
\usepackage{amssymb}
\newcommand{\be}{\begin{equation}}
\newcommand{\ee}{\end{equation}}
\newcommand{\bea}{\begin{eqnarray}}
\newcommand{\eea}{\end{eqnarray}}
\newcommand{\bi}{\begin{itemize}}
\newcommand{\ei}{\end{itemize}}
\newcommand{\ben}{\begin{enumerate}}
\newcommand{\een}{\end{enumerate}}
\newcommand{\bt}{\begin{tabbing}}
\newcommand{\et}{\end{tabbing}}

\newcommand{\pp}{{p^\prime}}
\newcommand{\bfp}{{\bf p}}
\newcommand{\bfpp}{{{\bf p}^\prime}}

\newcommand{\bfz}{{\bf 0}}
\newcommand{\dt}{{\Delta t}}
\newcommand{\dtp}{{\Delta t^\prime}}

\newcommand{\vp}{v^\prime}

\newcommand{\bfppp}{{\bf p}_\perp^\prime}

\title{
   \begin{picture}(0,0)(0,0)%
   \put(350,75){\makebox(0,0)[l]{\textnormal{\normalsize KEK-CP-371}}}%
   \end{picture}
   $B\!\to\!D^{(*)}\ell\nu$ form factors
   from lattice QCD with relativistic heavy quarks
}

\ShortTitle{
   $B\!\to\!D^{(*)}\ell\nu$ form factors
   from lattice QCD with relativistic heavy quarks
}

\author{
   JLQCD Collaboration:
   \speaker{T.~Kaneko}$^{a,b}$\thanks{E-mail: takashi.kaneko@kek.jp},
   Y.~Aoki$^c$,
   G.~Bailas$^a$,
   B.~Colquhoun$^d$,
   H.~Fukaya$^e$,
   S.~Hashimoto$^{a,b}$
   J.~Koponen$^{a}$
   \\
   \\
   \\
   \llap{$^a$}
   High Energy Accelerator Research Organization (KEK),
   Ibaraki 305-0801, Japan 
   \\
   \llap{$^b$}
   School of High Energy Accelerator Science,
   SOKENDAI (The Graduate University for Advanced Studies),
   Ibaraki 305-0801, Japan
   \\
   \llap{$^c$}
   RIKEN Center for Computational Science,
   Kobe 650-0047, Japan
   \\
   \llap{$^d$}
   Department of Physics and Astronomy,
   York University, Toronto, ON,
   M3J 1P3, Canada
   \\
   \llap{$^e$}
   Department of Physics, Osaka University, 
   Osaka 560-0043, Japan
}

\abstract{
We report on our calculation of the $B \to D^{(*)}\ell\nu$ form factors
in 2+1 flavor lattice QCD. The M\"obius domain-wall action is employed
for light, strange, charm and bottom quarks. At lattice cutoffs
$a^{-1} \sim 2.4$, 3.6 and 4.5 GeV, we simulate bottom quark masses
up to 0.7 $a^{-1}$ to control discretization errors. The pion mass
is as low as 230 MeV. We extrapolate the form factors to the continuum limit
and physical quark masses, and
make a comparison with recent phenomenological analyses.
}

\FullConference{
  37th International Symposium on Lattice Field Theory - Lattice2019\\
  16-22 June 2019\\
  Wuhan, China
}

\begin{document}

%// introduction ===============================================================

\section{Introduction}\label{sec:intro}

%// general background

The $B \to D^{(*)}\ell\nu$ semileptonic decays are promising probes
of new physics.
% For instance, more than 3~$\sigma$ tension with the Standard Model
% has been reported for the decay rate ratio
% $\Gamma(B\!\to\!D^{(*)}\tau\nu)/\Gamma(B\!\to\!D^{(*)}\{e,\mu\}\nu)$
% for the lepton flavor universality violation~\cite{HFLAV}.
However, there has been a long-standing tension
in the Cabibbo-Kobayashi-Maskawa (CKM) matrix element $|V_{cb}|$
with its alternative determination from inclusive decays~\cite{HFLAV}.
Phenomenological analyses~\cite{Vcb:BLPR,Vcb:BGS,Vcb:GK,Vcb:BGS:2,Vcb:BLPR:2,Vcb:GJS}
of recent Belle data of the $B\!\to\!D^*\ell\nu$ differential decay rate~\cite{B2D*:exprt:Belle:tag:unfold,B2D*:exprt:Belle:untag:unfold}
are deepening our understanding of the systematics of
the $|V_{cb}|$ determination.
An unambiguous resolution of the $|V_{cb}|$ tension, however,
will require a first-principle calculation of the relevant form factors
by means of lattice QCD. 
The $B\!\to\!D^*\ell\nu$ form factors at non-zero recoils
are of particular importance, and are being calculated
by us and other collaborations~\cite{B2D*:Fermilab/MILC:lat19,Review:B2D*:lat19}.

%// JLQCD's study

In this article,
we update our results for the $B\!\to\!D^{(*)}\ell\nu$ form factors.
After our previous report~\cite{B2D*:Nf2+1:JLQCD:lat18},
the calculation has been extended to a larger cutoff $a^{-1}\!\sim\!4.5$~GeV
and a smaller pion mass $M_{\pi}\!\sim\!230$~MeV.
A notable feature of our simulations is the use of 
relativistic quark formulation with good chiral symmetry
for all the relevant flavors.
This enables us to straightforwardly study
interesting $B$ meson decays
including $B\!\to\!\ell\nu$~\cite{fB:Nf2+1:JLQCD:Lat16}
and inclusive decay~\cite{Incl:Nf2+1:JLQCD:lat18}.
Our studies of 
$B\!\to\!\pi\ell\nu$~\cite{B2pi:Nf2+1:JLQCD:lat19},
$B\!\to\!D^{**}\ell\nu$~\cite{B2D**:Nf2+1:JLQCD:lat19}
and 
$B\!\to\!K\ell\ell$~\cite{B2Kll:Nf2+1:JLQCD:lat19}
are also reported in these proceedings.

%// simulation =================================================================

\section{Calculation of form factors}\label{sec:ff}

\begin{table}[b]
\centering
\small
% \begin{flushleft}
\caption{
  Simulation parameters. Quark masses are bare value in lattice units.
  % The rightmost column shows the number of
  % temporal locations of the meson source operator
  % of correlation functions.
}
% \end{flushleft}
% \vspace{0mm}
\label{tbl:sim:param}
\begin{tabular}{l|llll}
   \hline 
   lattice parameters 
   & $m_{ud}$ & $m_s$ & $M_\pi$[MeV] & $M_K$[MeV] % & $N_{\dt + \dtp}$
   \\ \hline
   $\beta\!=\!4.17$, \ 
   $a^{-1}\!=\!2.453(4)$, \ 
   $32^3\!\times\!64\!\times\!12$
   & 0.0190 & 0.0400 & 499(1) & 618(1) % & 4
   \\
   & 0.0120 & 0.0400 & 399(1) & 577(1) % & 4
   \\
   & 0.0070 & 0.0400 & 309(1) & 547(1) % & 4
   \\ \hline
   $\beta\!=\!4.17$, \ 
   % $a^{-1}\!=\!2.453(4)$, \ 
   $48^3\!\times\!96\!\times\!12$
   & 0.0035 & 0.0400 & 226(1) & 525(1) % & 2
   \\ \hline
   $\beta\!=\!4.35$, \ 
   $a^{-1}\!=\!3.610(9)$, \ 
   $48^3\!\times\!96\!\times\!8$
   & 0.0120 & 0.0250 & 501(2) & 620(2) % & 4
   \\
   & 0.0080 & 0.0250 & 408(2) & 582(2) % & 4
   \\
   & 0.0042 & 0.0250 & 300(1) & 547(2) % & 4
   \\ \hline
   $\beta\!=\!4.47$, \ 
   $a^{-1}\!=\!4.496(9)$, \ 
   $64^3\!\times\!128\!\times\!8$
   & 0.0030 & 0.0150 & 284(1) & 486(1) % & 2
   \\ \hline
\end{tabular}
% \vspace{0mm}
\end{table}

We generate gauge ensembles of 2+1 flavor lattice QCD
at cutoffs of 2.5\,--\,4.5~GeV.
Chiral symmetry is preserved to good accuracy
by employing % the tree-level improved Symanzik gauge action,
the M\"obius domain-wall quark action~\cite{MDWF,MDWF:JLQCD:lat13}.
This simplifies the renormalization of the relevant matrix elements,
which often suffers from large discretization errors. 
We simulate a strange quark mass $m_s$ close to its physical value,
whereas the degenerate up and down quark mass $m_{ud}$
corresponds to pion masses as low as $M_\pi\!\sim\!230$~MeV.
The spatial lattice size $L$ is chosen to satisfy
a condition $M_\pi L \!\gtrsim\!4$ to control finite volume effects.
The statistics are 5,000 Molecular Dynamics time
at each simulation point.
These simulation parameters are listed in Table~\ref{tbl:sim:param}.

%// measurement 

The $B\!\to\!D^{(*)}$ matrix elements are parametrized
by six form factors in total,
\bea
   \sqrt{M_B M_D}^{-1}
   \langle D(\pp) | V_\mu | B(p) \rangle
   & = &
   (v+\vp)_\mu h_+(w) + (v-\vp)_\mu h_-(w),
   \label{eqn:ff:B2D}
   \\[2mm]
   \sqrt{ M_B M_{D^*} }^{-1}
   \langle D^*(\epsilon,\pp) | V_\mu | B(p) \rangle
   & = &
   \varepsilon_{\mu\nu\rho\sigma} \, \epsilon^{*\nu} v^{\prime\rho} v^\sigma \, h_V(w),
   \label{eqn:ff:B2D*:V}
   % \eea
   % \bea
   \\[2mm]
   \sqrt{ M_B M_{D^*} }^{-1}
   \langle D^*(\epsilon,\pp) | A_\mu | B(p) \rangle
   & = &
   -i(w+1) \, \epsilon_\mu^* \,   h_{A_1}(w)
   + i(\epsilon^* v)\, v_\mu \, h_{A_2}(w) + i(\epsilon^* v)\, \vp_\mu \, h_{A_3}(w),
   \hspace{10mm}
   \label{eqn:ff:B2D*:A}
\eea
where
$w=v\vp$ is the recoil parameter
defined by four velocities $v\!=\!p/M_B$ and $\vp\!=\!\pp/M_{D^{(*)}}$,
and $\epsilon$ is the polarization vector of $D^*$,
which satisfies $\pp\epsilon\!=\!0$.

%// 3pt func. 

We employ the M\"obius domain-wall action also for charm and bottom quarks
to calculate $B\!\to\!D^*$ three-point functions.
The charm quark mass $m_c$ is set to its physical value
determined from the spin averaged mass
$(M_{\eta_c}+3M_{J/\Psi})/4$, % low-lying charmonium spectrum,
whereas we use the bottom masses 
$m_b\!=\!1.25 m_c, 1.25^2 m_c, ...$
under a condition $m_b\!<\!0.7\,a^{-1}$
to suppress discretization effects.
The $B\!\to\!D^{(*)}$ matrix elements can be extracted from the
three-point functions, provided that they are dominated by
their ground state contribution as 
\bea
   % &&
   C_{\mathcal O_\Gamma}^{BD^{(*)}}(\dt,\dtp;\bfp,\bfpp)
   % \nn \\
   % & = & 
   % \frac{1}{N_{\tsrc}}\sum_{\tsrc}
   % \sum_{\bfxsrc,\bfx,\bfxp}
   % \langle 
   %    {\mathcal O}_{D^{(*)}}(\bfxp,\tsrc+\dt+\dtp)
   %    {\mathcal O}_\Gamma(\bfx,\tsrc+\dt)
   %    {\mathcal O}_{B}(\bfxsrc,\tsrc)^\dagger
   % \rangle
   % e^{-i\bfp(\bfx-\bfxsrc)-i\bfpp(\bfxp-\bfx)}
   % \nn \\
   % & \to &
   & \xrightarrow[\dt,\dtp \to \infty ]{} &
   \frac{Z_{D^{(*)}}^*(\bfpp)\,Z_B(\bfp)}{4E_{D^{(*)}}(\bfpp)E_B(\bfp)}
   \langle D^{(*)}(\pp) | {\mathcal O}_\Gamma | B(p) \rangle
   e^{-E_{D^{(*)}}(\bfpp)\dtp -E_B(\bfp)\dt },
   % \hspace{3mm} (\dt, \dtp \to \infty),
   \label{eqn:ff:corr_3pt}
\eea
where ${\mathcal O}_\Gamma\!=\!V_\mu$ or $A_\mu$,
and the argument $\epsilon$ is suppressed for $Z_{D^*}$
and $|D^*(\pp)\rangle$.
We apply the Gaussian smearing to the interpolating field
${\mathcal O}_P$ ($P\!=\!B,D,D^*$) to enhance its overlap
to the ground state
$Z_P(\bfp)\!=\!\langle P(p) | {\mathcal O}_P^\dagger \rangle$.
The $B$ meson is at rest ($\bfp\!=\!\bfz$),
and the $w$ dependence of the form factors
is studied by varying the three momentum of $D^{(*)}$ as 
$|\bfpp|^2\!=\!0,1,2,3,4$ in units of $(2\pi/L)^2$.

%// improvement of statistical accruacy: ratios

For precise determination of the form factors,
we construct ratios of the correlation functions,
in which unnecessary overlap factors and exponential damping factors cancel~\cite{double_ratio}. The statistical fluctuation is also expected to partly cancel.
For instance, the normalizations, $h_+(1)$ and $h_{A_1}(1)$,
and a ratio $R_1(w)\!=\!h_V(w)/h_{A_1}(w)$
can be directly extracted from the ratios
\bea
   R_{(k)}^{BD^{(*)}}(\dt,\dtp)
   & = &
   \frac{ C_{V_4(A_k)}^{BD^{(*)}}(\dt,\dtp;\bfz,\bfz)\,
          C_{V_4(A_k)}^{D^{(*)}B}(\dt,\dtp;\bfz,\bfz) }
        { C_{V_4(A_k)}^{BB}(\dt,\dtp;\bfz,\bfz)\,
          C_{V_4(A_k)}^{D^{(*)}D^{(*)}}(\dt,\dtp;\bfz,\bfz) }
   \xrightarrow[\dt,\dtp\to\infty]{}
   |h_{+(A_1)}(1)|^2,
   \label{eqn:ff:b2d+b2d*:r1}
   \\[-1mm]
   R_{V,k}^{BD^*}(\dt,\dtp;\bfz,\bfppp)
   & = &
   \frac{ C_{V_k}^{BD^*}(\dt,\dtp;\bfz,\bfppp) }
        { C_{A_k}^{BD^*}(\dt,\dtp;\bfz,\bfppp) }
   \xrightarrow[\dt,\dtp\to\infty]{}
   \frac{ \epsilon_{kij} \epsilon^*_i v_{\perp j}^\prime}{1+w}
   \frac{h_V(w)}{h_A(w)},
   \label{eqn:ff:b2d*:r3}
\eea
where $\bfppp$ represents the $D^*$ momentum satisfying $v \epsilon\!=\! 0$.
% These are important imput 
% in the conventional determination of $|V_{cb}|$.
We refer the readers to Refs.~\cite{B2D*:Nf2+1:JLQCD:lat18,B2D*:Fermilab/MILC:lat17} for ratios to determine other form factors.

\FIGURE{
   \label{fig:ff:drat14}
   \includegraphics[angle=0,width=0.48\linewidth,clip]%
                   {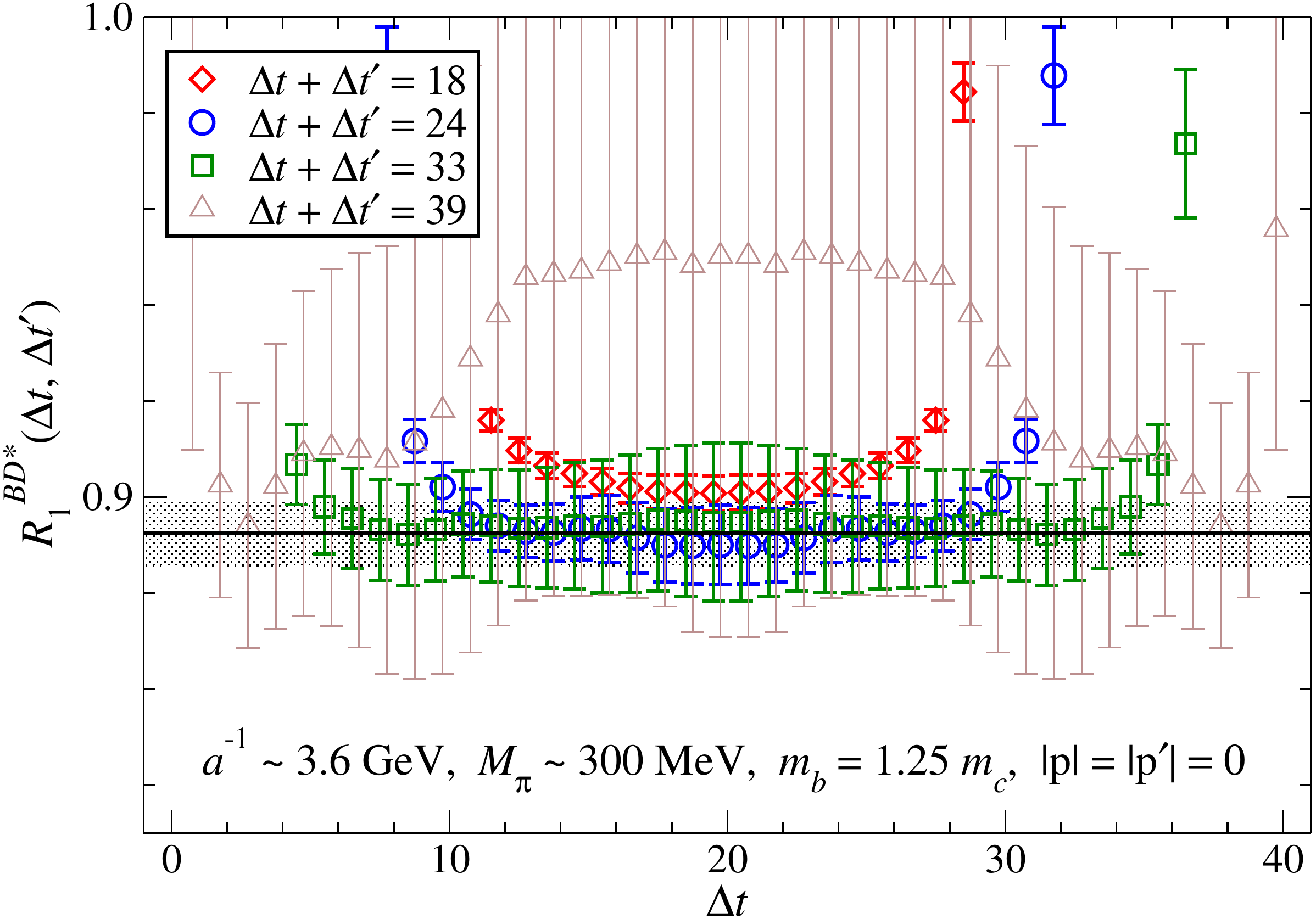}
   %\vspace{0mm}
   \caption{
      Double ratio~(\ref{eqn:ff:b2d+b2d*:r1})
      % $R_1^{BD^{(*)}}(\dt,\dtp)$
      as a function of the temporal location $\Delta t$ of $A_\mu$.
      We plot data at $\beta\!=\!4.35$, $m_{ud}\!=\!0.0042$
      and $m_b\!=\!1.25\,m_c$.
      Symbols with different shapes show data with different values
      of $\dt+\dtp$.
      The shaded band shows a constant fit to estimate $|h_{A_1}(1)|^2$.
   }
   \vspace{-5mm}
}
A salient feature of this analysis is that
the form factors in Eqs.~(\ref{eqn:ff:B2D})\,--\,(\ref{eqn:ff:B2D*:A})
can be calculated without finite renormalization of
the local lattice operator
${\mathcal O}_\Gamma\!=\!V_\mu, A_\mu$ in Eq.~(\ref{eqn:ff:corr_3pt}).
The relevant renormalization factors cancel in the correlator ratios
with the relativistic heavy quarks with chiral symmetry.
This is an advantage toward a precision calculation of the form factors,
because we observe large discretization effects
to the wave-function renormalization factor~\cite{fB:Nf2+1:JLQCD:Lat16}.

%// multiple 

The choice of the source-sink separation $\dt+\dtp$ 
in Eq.~(\ref{eqn:ff:corr_3pt}) is also crucial
for the precision study of heavy hadrons~\cite{Review:HQPhys:Hashimoto}.
Except for the on-going simulation at the largest $a^{-1}$
and smallest $M_\pi$,
we repeat our measurement for four different values of
$\dt+\dtp$ in a range 0.7\,--\,2.2~fm.
Figure~\ref{fig:ff:drat14},
which shows the double ratio $R_{1}^{BD^*}(\dt,\dtp)$, demonstrates that,
towards larger separation,
the three-point function has less excited state contamination,
but its statistical noise rapidly grows.
With the four values of $\dt+\dtp$,
we can safely identify the plateau corresponding to the ground state dominance.
The statistical accuracy is typically 1\,--\,2\% for
$h_+$, $h_{A_1}$ and $h_V$.
Other form factors are less accurate,
because i) $h_-$ and $h_{A_2}$ are close to zero due to heavy quark symmetry,
and ii) we do not have matrix element, which is exclusively sensitive
to $h_{A_2}$ or $h_{A_3}$, partly due to our kinematical setup
with the $B$ meson at rest.

%// continuum chiral extrapolation =============================================

\section{Continuum and chiral extrapolation}\label{sec:ccfit}

In this preliminary report,
we extrapolate the form factors to the continuum limit
and physical quark masses by using the following simple form
based on the next-to-leading order (NLO) 
heavy meson chiral perturbation theory (HMChPT)~\cite{B2D*:HMChPT:RW,B2D*:HMChPT:S}
\bea
   h_X
   & = &
   c + F_{\rm log}(M_\pi,f,\Lambda_\chi,g,\Delta_c)
   + c_w (w-1) + \frac{c_b}{m_b} + c_\pi M_\pi^2 + c_{\eta_s} M_{\eta_s}^2 + c_{a} a^2
   + d_w (w-1)^2,
   \label{eqn:ccfit:form}
\eea   
where $M_{\eta_s}^2\!=\!2M_K^2-M_\pi^2\!\sim\!2m_s$ is used
to describe the (presumably small) $m_s$ dependence.
The chiral log of, for instance, $h_{A_1}$ is given as
\bea
   F_{\rm log}(M_\pi,f,\Lambda_\chi,g,\Delta_c)
   & = &
   \frac{g^2}{32 \pi^2 f^2} \bar{F}(M_\pi,\Delta_c,\Lambda_\chi).
   \label{eqn:ccfit:chirallog}
\eea
We refer the readers to Refs.~\cite{B2D*:HMChPT:RW,B2D*:HMChPT:S}
for the exact form of the loop integral function
$\bar{F}\!=\!\Delta_c^2 \ln[ M_\pi^2 / \Lambda_\chi^2 ] + O(\Delta_c^3)$.
The decay constant in the chiral limit $f$ and
the $D^*$\,--\,$D$ mass splitting $\Delta_c$ are fixed to
the experimental value of the pion decay constant $f_\pi$
and $M_{D^*}\!-\!M_D$, respectively.
The renormalization scale of HMChPT is set to $\Lambda_\chi\!=\!4\pi f_\pi$.
These choices only modify the higher order chiral corrections.
The coupling $g$ is set to the value 0.53(8) 
quoted in Ref.~\cite{B2D*:Nf2+1:Fermilab/MILC:w1},
which covers the previous estimates of
the $D^*D\pi$, $B^*B\pi$ couplings and their static limit.
We note that, in this preliminary report,
the quoted error is statistical only.

\begin{figure}[tb] 
  \centering
  \includegraphics[width=0.49\linewidth,clip]{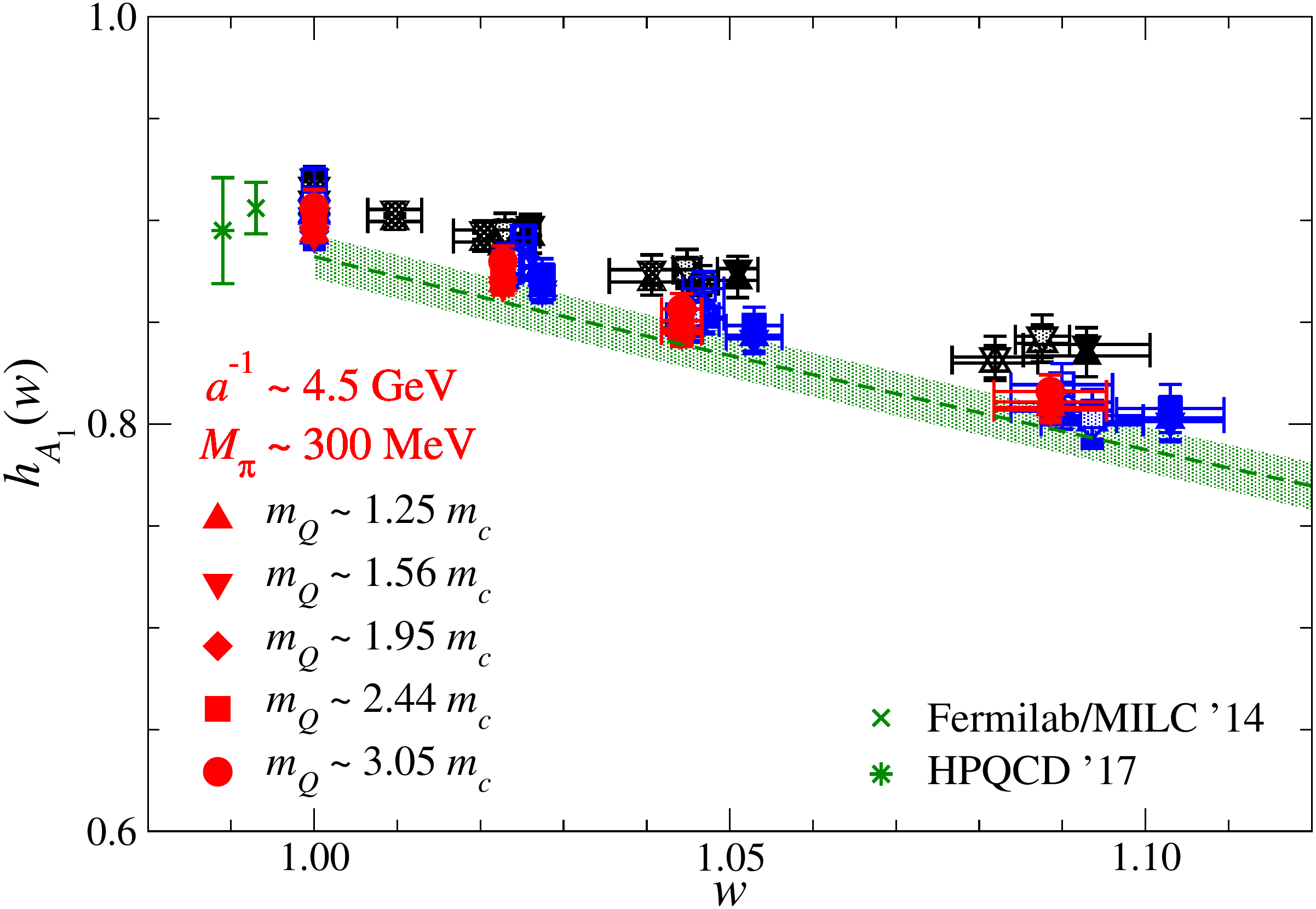}
  \hspace{1mm}
  \includegraphics[width=0.49\linewidth,clip]{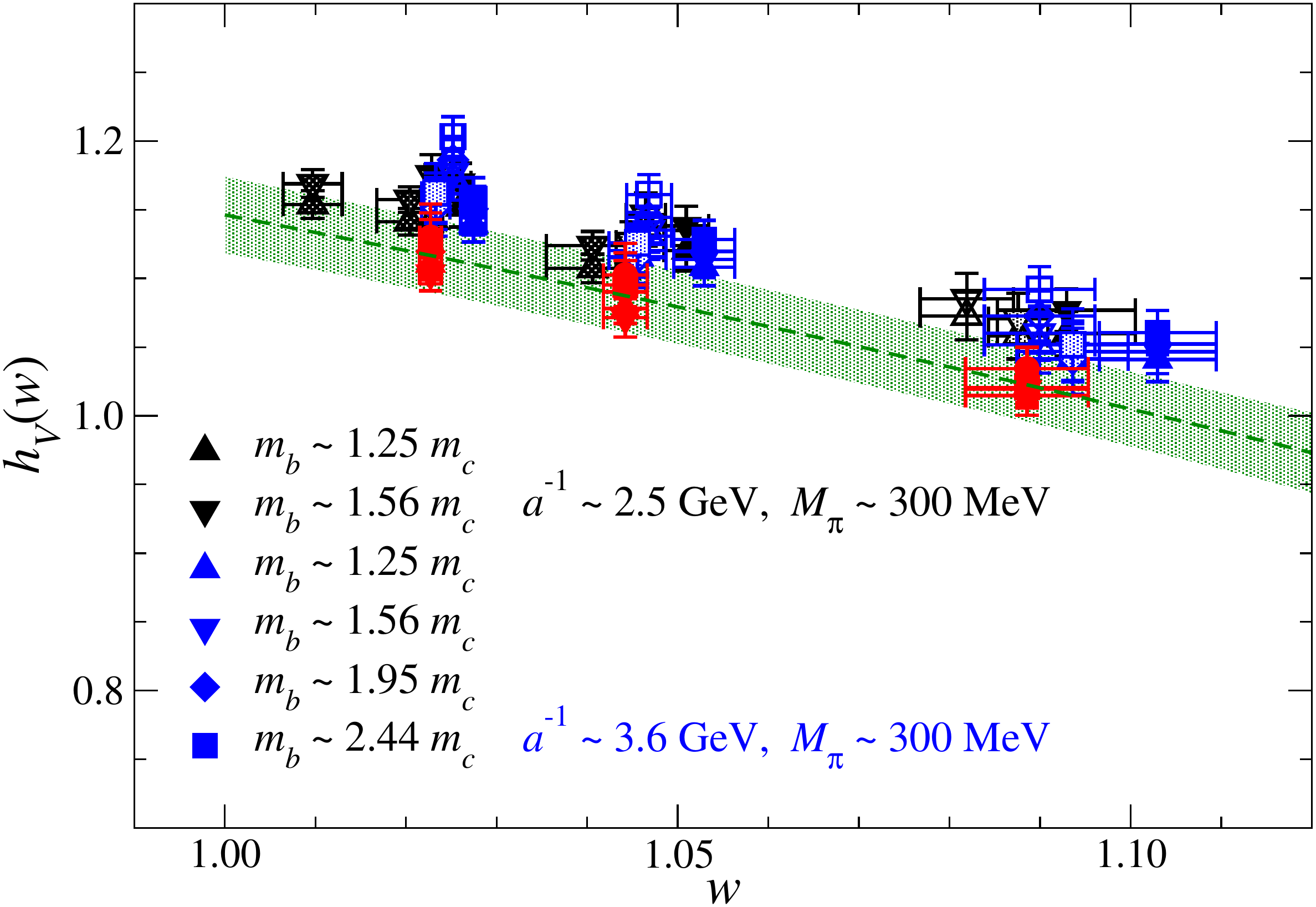}
  \vspace{-5mm}
  
  \includegraphics[width=0.49\linewidth,clip]{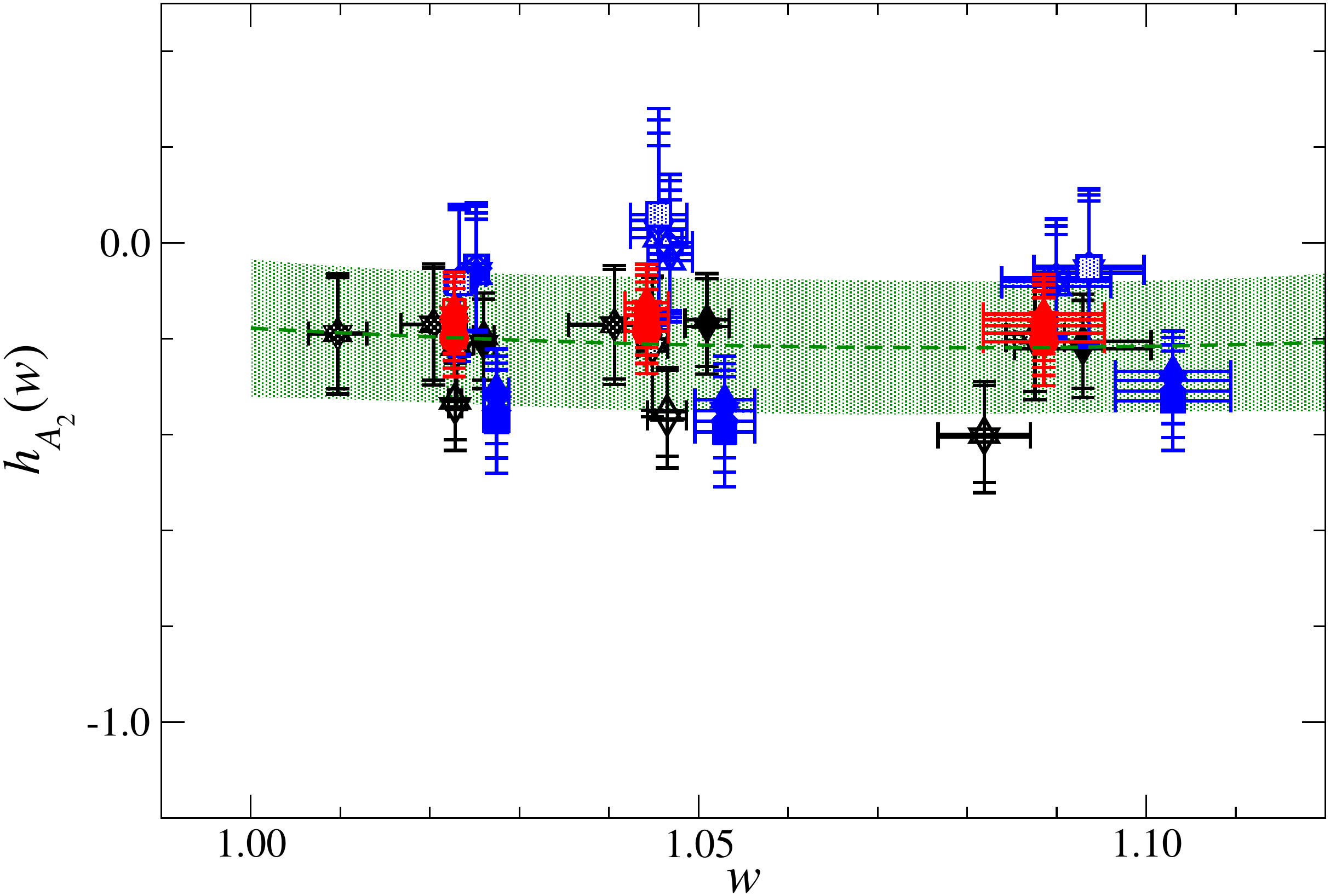}
  \hspace{1mm}
  \includegraphics[width=0.49\linewidth,clip]{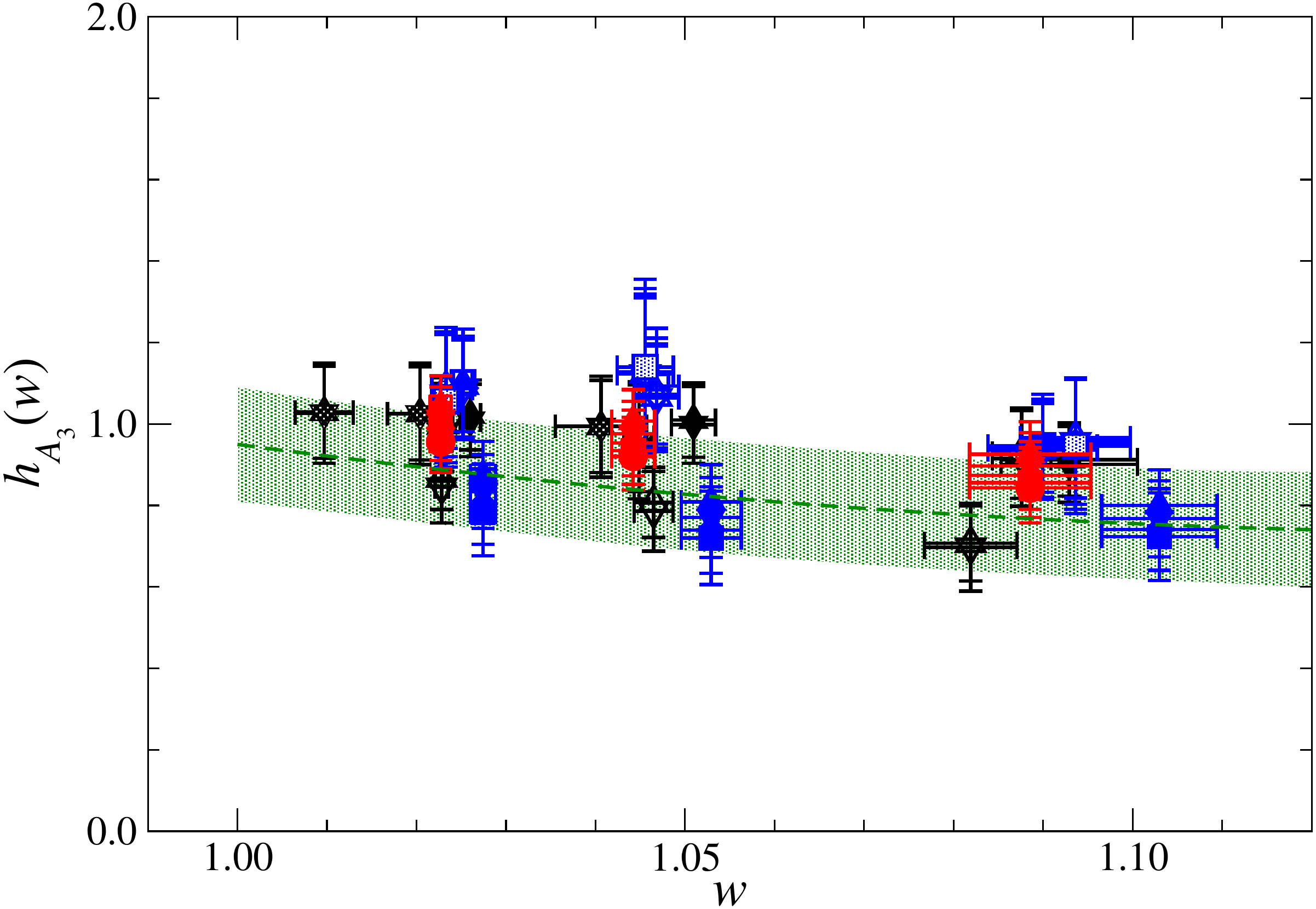}
  \vspace{-8mm}

  \caption{
    Form factors of the $B\!\to\!D^*\ell\nu$ decay as a function of $w$.
    Symbols show data at simulation points.
    Those extrapolated to the continuum limit and physical quark masses
    are plotted by the green bands.
    The red, blue, black symbols are obtained at $a^{-1}\!=\!4.5$, 3.6
    and 2.5~GeV,
    whereas open, pale shaded, filled, dark shaded symbols are
    at $M_\pi\!\sim\!500$, 400, 300 and 230~MeV, respectively.
    Symbols with different shapes show data with different values of $m_b$.
    For $h_{A_1}(1)$, we also plot the previous estimates~\cite{B2D*:Nf2+1:Fermilab/MILC:w1,B2D*:HPQCD:w1}.
  }
  \label{fig:ccfit:B2D*}
  \vspace{-3mm}
\end{figure}

Figure~\ref{fig:ccfit:B2D*} shows $B\!\to\!D^*\ell\nu$ form factors
at simulated points and those extrapolated to the continuum limit
and physical quark masses.
Our result for $h_{A_1}(1)$ is in reasonable agreement with
the previous estimate by Fermilab/MILC~\cite{B2D*:Nf2+1:Fermilab/MILC:w1}
and HPQCD~\cite{B2D*:HPQCD:w1}.
We observe a mild dependence of the form factors on $a^{-1}$ and quark masses,
and the $w$ dependence does not show any strong curvature
in our simulation region near $w\!=\!1$.
As a result,
many of fit parameters in Eq.~(\ref{eqn:ccfit:form}) turned out
to be consistent with zero.
Only $c$, $c_w$ and $c_b$ for $h_+$, $h_{A_1}$ and $h_V$
have a statistical error less than 50\,\%.
Since the parameter dependences are described reasonably well
by a constant or linear term,
this continuum and chiral extrapolation may not suffer from 
large systematic uncertainties, which are under investigation.

%// comparison =================================================================

\section{Implication to $|V_{cb}|$ determination}

%// theory background

In the limit of $m_\ell\!=\!0$,
the $B\!\to\!D^*\ell\nu$ differential decay rate is described
by $h_{A_1}(w)$ and two ratios $R_1(w)$ (defined above)
and $R_2(w)\!=\!(r h_{A_2}+ h_{A_3})/h_{A_1}$ ($r\!=\!M_{D^*}/M_B$) as
\bea
   \frac{d\Gamma}{dw}
   \propto 
   \frac{G_F^2 |V_{cb}|^2}{48\pi^3} 
   \left[
      2\frac{1-2wr+r^2}{(1-r)^2}
     \left\{ 1 + \frac{w-1}{w+1} R_1(w)^2 \right\}
   +\left\{ 1 + \frac{w-1}{1-r}\left( 1-R_2(w) \right) \right\}^2
   \right]
   h_{A_1}(w)^2.
   \hspace{5mm}
   \label{eqn:comp:dGamma_dw}
\eea     
The conventional determination of $|V_{cb}|$
employs the Caprini-Lellouch-Neubert (CLN) parametrization~\cite{B2D*:FF:CLN},
in which $h_{A_1}$, $R_1$ and $R_2$ are expanded in terms of 
a small kinematical parameter
% $z\!=\!(\sqrt{w+1}-\sqrt{2}a)/(\sqrt{w+1}+\sqrt{2}a)$,
% where $a$ is a tunable parameter,
and some of expansion coefficients are constrained
by heavy quark effective theory (HQET) supplemented by the QCD sum rule inputs.
Recent Belle data with the full kinematical distribution, on the other hand,
enables an analysis with the Boyd-Grinstein-Lebed (BGL) parametrization
without such HQET constraints and hence involving more free parameters.

%// R1

\begin{figure}[tb] 
  \centering
  \includegraphics[width=0.49\linewidth,clip]{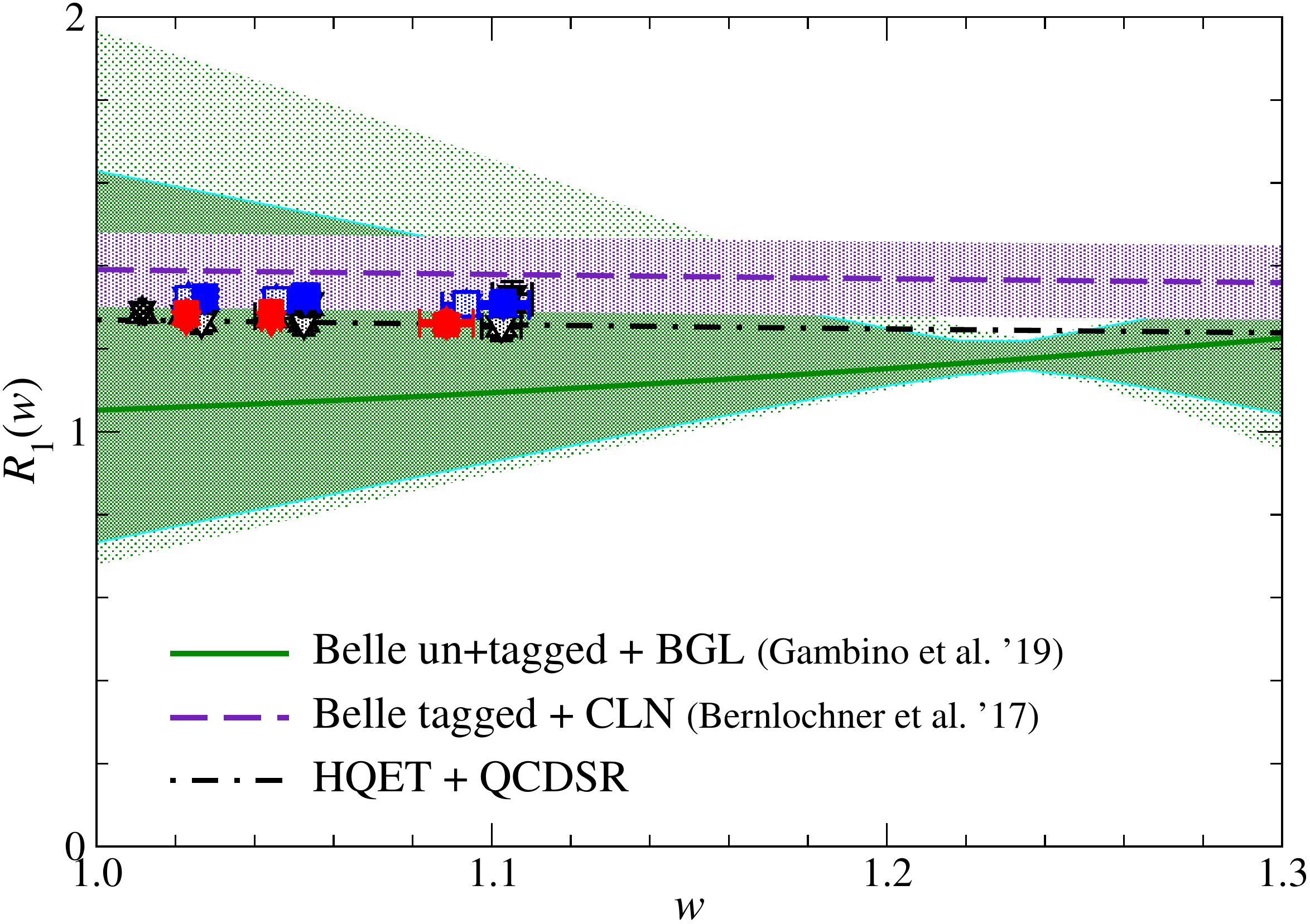}
  \hspace{1mm}
  \includegraphics[width=0.49\linewidth,clip]{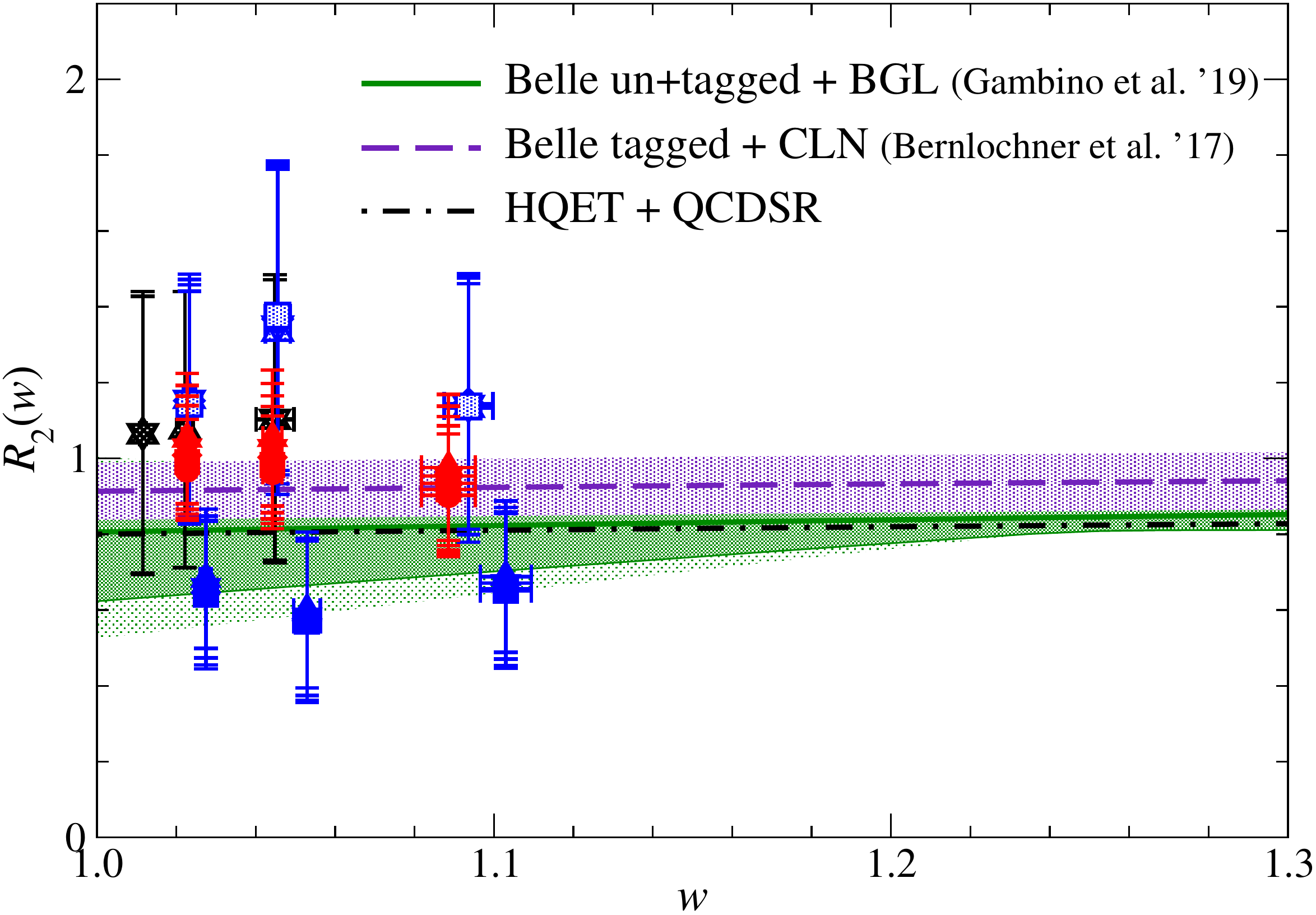}
  \vspace{-8mm}
  
  \caption{
    Form factor ratios $R_1$ (left panel) and $R_2$ (right panel)
    as a function of $w$.
    The symbols show our data at simulation points.
    The pale and dark shaded green bands show the results
    of the recent BGL fits
    with the standard and strong unitarity bounds~\cite{Vcb:GJS},
    whereas the purple band is from the CLN fit~\cite{Vcb:BLPR:2}.
    We also plot the NLO HQET prediction by the dot-dashed line.
  }
  \label{fig:comp:R12}
  \vspace{-3mm}
\end{figure}

It was reported a few years ago
that i) model independent fit of the Belle tagged data~\cite{B2D*:exprt:Belle:tag:unfold}
with the BGL parametrization yielded
$|V_{cb}|$ consistent with the inclusive determination~\cite{Vcb:BGS,Vcb:GK},
and that ii) there was a clear difference in $R_1$ between the BGL and CLN fits~\cite{Vcb:BLPR:2}.
At last year's conference, we reported that
our lattice data favor $R_1$ from the CLN fit.
This is further confirmed by the additional data
at the largest $a^{-1}$ and smallest $M_\pi$
as shown in the left panel of Fig.~\ref{fig:comp:R12}.
Meanwhile, the BGL fit has been updated
by including the Belle untagged data~\cite{B2D*:exprt:Belle:untag:unfold},
and discrepancy from the CLN fit and lattice QCD has been resolved~\cite{Vcb:GJS}.

%// R2

\begin{figure}[b] 
  \centering
  \includegraphics[angle=0,width=0.48\linewidth,clip]%
                  {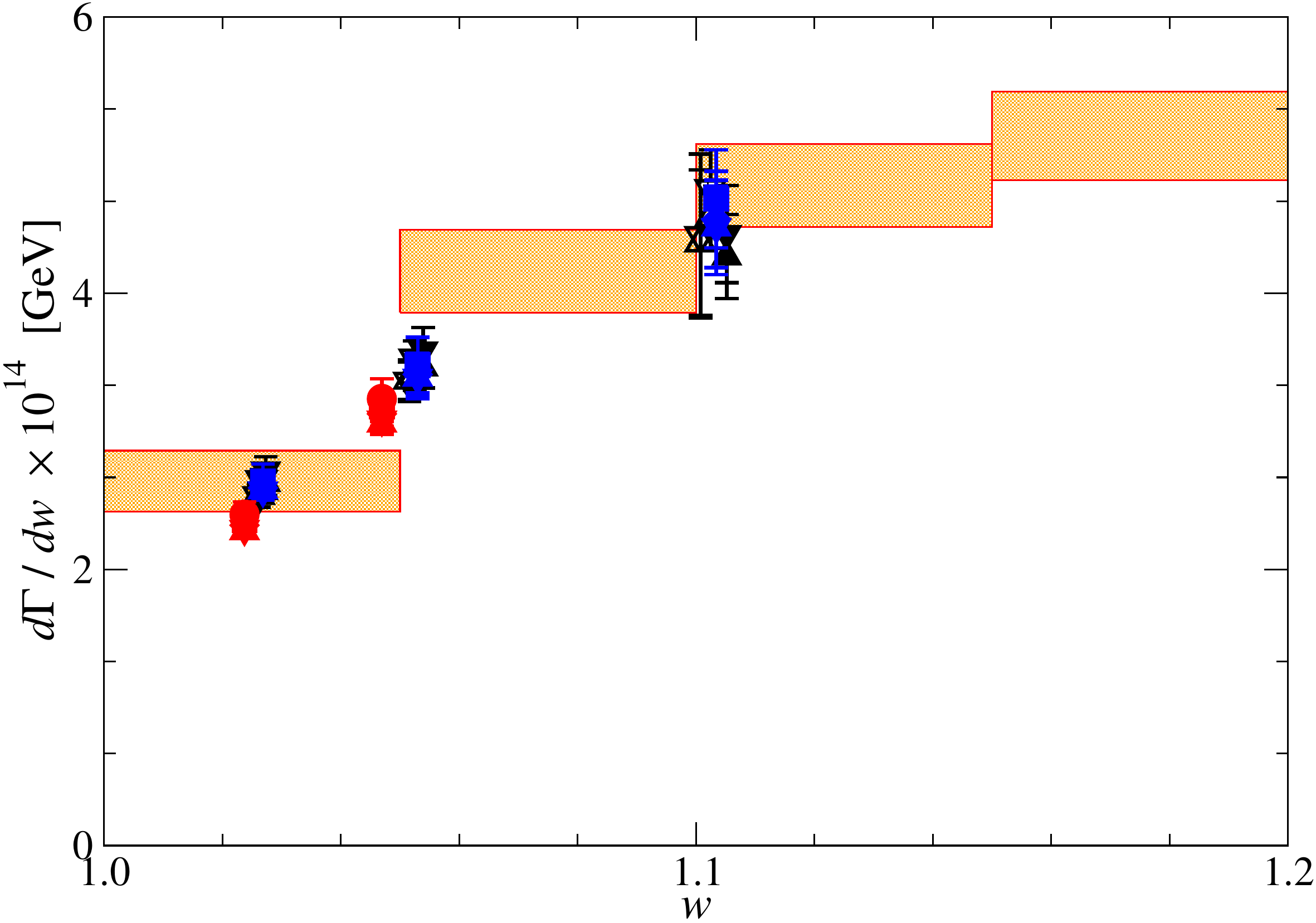}
  \hspace{1mm}
  \includegraphics[angle=0,width=0.48\linewidth,clip]%
                  {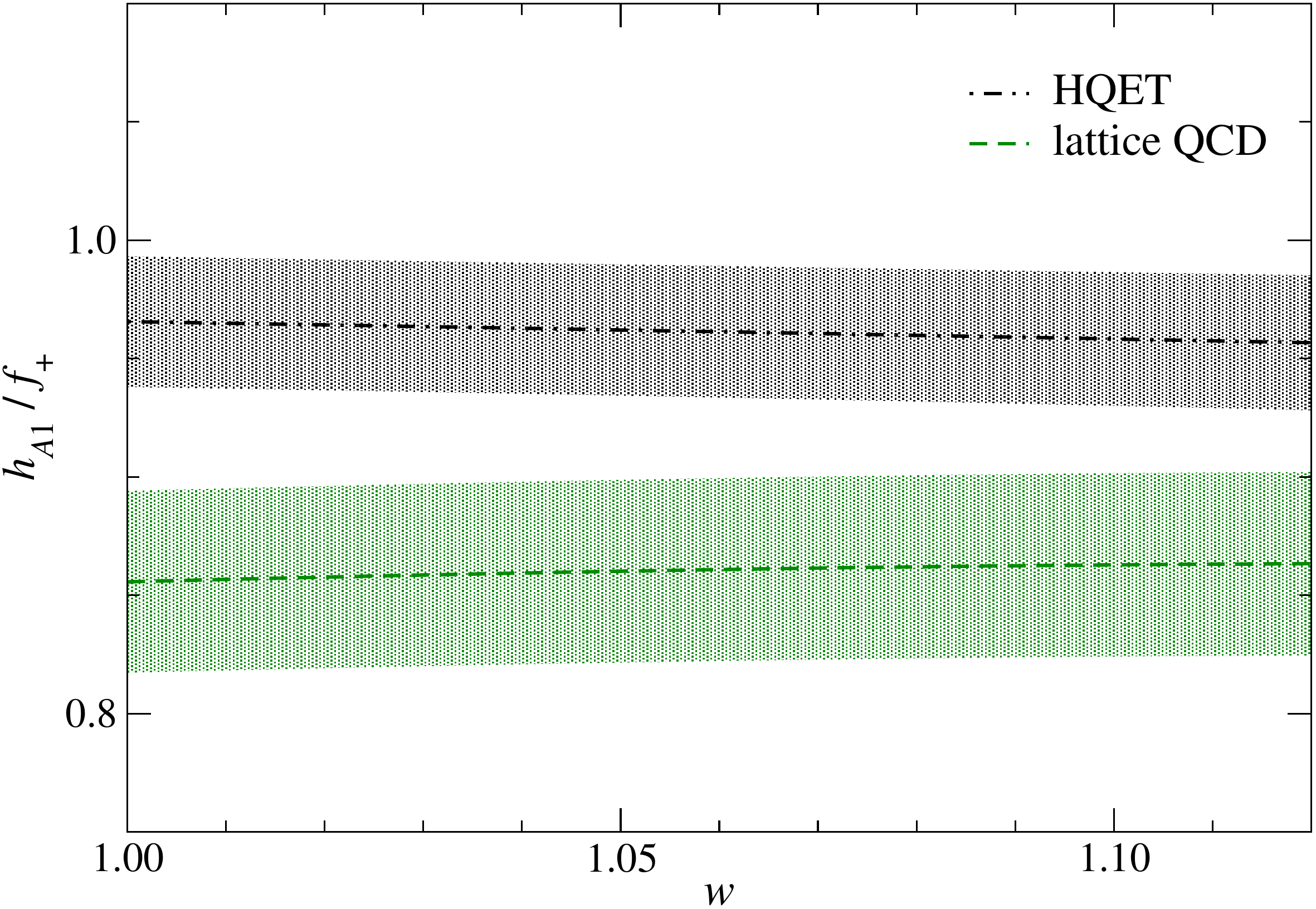}
  \vspace{-3mm}
  \caption{
    Left panel:
    $B\!\to\!D^*\ell\nu$ differential decay rate $d\Gamma/dw$
    as a function of $w$. Symbols are estimated from our data
    at simulation points, whereas the orange band shows
    Belle data~\cite{B2D*:exprt:Belle:tag:unfold}.
    We assume $|V_{cb}|$ from $B\!\to\!D^*\ell\nu$~\cite{HFLAV}
    to estimate $d\Gamma/dw$.
    Right panel:
    $h_{A_1}/f_+$ as a function of $w$.
    Our lattice result and the NLO HQET prediction are
    plotted by the green and black bands,respectively.
  }
  \label{fig:comp:dGdw+hA1V1}
  \vspace{-8mm}
\end{figure}

On the other hand, there has been no large difference in $R_2$
between the BGL and CLN fits
as shown in the right panel of Fig.~\ref{fig:comp:R12}.
Our lattice data are consistent with these phenomenological estimates
within relatively large uncertainty coming from $h_{A_2}$ and $h_{A_3}$.
We note that this uncertainty is not problematic in predicting
the differential decay rate~(\ref{eqn:comp:dGamma_dw}),
because the contribution of $R_2$ is suppressed by a factor $w-1$
in our simulation region near $w\!=\!1$.
The left panel of Fig.~\ref{fig:comp:dGdw+hA1V1} demonstrates that 
we can estimate $d\Gamma/dw$ with an accuracy comparable to experiments,
and also shows a reasonable agreement between our and experimental data.

%// hA1/f+

A ratio $h_{A_1}/f_+$,
where $f_+$ is the vector form factor for $B\!\to\!D\ell\nu$,
is also an important quantity,
since the CLN parametrization of $h_{A_1}$
is derived from this ratio in NLO HQET
and a dispersive parametrization of $f_+$~\cite{B2D*:FF:CLN}.
The right panel of Fig.~\ref{fig:comp:dGdw+hA1V1}
shows a reasonable agreement in the $w$ dependence
between HQET and lattice QCD.
While there is a $\sim\!10$~\% difference in the normalization,
this does not necessarily lead to the $|V_{cb}|$ tension,
since $h_{A_1}(1)$ is absorbed into the overall factor
of $d\Gamma/dw$, which is treated as a fit parameter
in the $|V_{cb}|$ determination.

%// conclusion =================================================================

\section{Summary}

In this article,
we report on our studies of the $B\!\to\!D^{(*)}\ell\nu$ decays.
The relevant form factors are precisely determined
by simulating multiple values of the source-sink separation.
While the systematics of the continuum and chiral extrapolation
are under investigation, it is expected to be reasonably controllable
due to the mild parametric dependence of the form factors.

Recently, it is argued that
the constraint in the CLN parametrization
is responsible for the $|V_{cb}|$ tension.
Except the normalization of $h_{A_1}/f_+$,
our lattice data of $h_{A_1}/f_+$, $R_1$ and $R_2$
show a reasonable agreement with the CLN fit and NLO HQET.
A more detailed analysis,
such as the CLN and BGL fits both to lattice and experimental data,
is needed towards an unambiguous resolution of the $|V_{cb}|$ tension.

%// Acknowledgment =============================================================

We are grateful for F.U.~Bernlochner, Z.~Ligeti, M.~Papucci, D.J.~Robinson,
and P.~Gambino, M.~Jung, S.~Schacht 
for making their numerical results in Refs.~\cite{Vcb:BLPR:2,Vcb:GJS}
available to us.
Numerical simulations are performed on Oakforest-PACS
at JCAHPC under a support of the HPCI System Research Projects
(Project ID: hp180132 and hp190118)
and Multidisciplinary Cooperative Research Program
in CCS, University of Tsukuba
(Project ID: xg18i016).
This work is supported in part by JSPS KAKENHI Grant Number JP18H03710.
% 16K05320, 18H01216, 18H03710 and 18H04484. 

%// references =================================================================

\end{document}